\documentclass[preprint,showkeys,superscriptaddress,amsmath,amssymb]{revtex4-1}

\usepackage{newtxtext,newtxmath}
\usepackage[T1]{fontenc}
\usepackage{graphicx}
\usepackage{bm}
\usepackage{textcomp}
\usepackage{braket}
\usepackage{siunitx}
\usepackage{enumitem}
\usepackage{caption}
\DeclareCaptionLabelSeparator{bar}{ $|$ }
\usepackage[colorlinks=true,
linkcolor=magenta,
citecolor=blue,
pdfauthor={Hyounghan Kwon},
]{hyperref}

\usepackage{times}
\usepackage{dcolumn}
\usepackage{amsmath}
\usepackage[dvipsnames]{xcolor}
\usepackage[labelsep=space]{caption}
\usepackage{setspace}
\usepackage{amssymb}
\usepackage{amsmath}
\usepackage[version=3]{mhchem}
\usepackage{lineno}

\usepackage{booktabs}
\usepackage{array}
\usepackage{makecell}
\usepackage{tabularx}
\usepackage[table]{xcolor}

\usepackage[percent]{overpic}
\usepackage{xcolor}

\newcommand{\NatureFormat}{%
		\renewcommand{\figurename}{\textbf{Fig.}}
     }
\NatureFormat

\begin{document}

\title{Nonlinear Electro-Optic Visible Photonic Circuits for Solid-State Quantum Defects}

\author{Yongchan Park}
\thanks{These authors contributed equally}
\affiliation{Center for Quantum Technology, Korea Institute of Science and Technology (KIST), Seoul 02792, South Korea}
\affiliation{School of Electrical Engineering, Korea University, Seoul 02841, South Korea}

\author{Yong Soo Lee}
\thanks{These authors contributed equally}
\affiliation{Center for Quantum Technology, Korea Institute of Science and Technology (KIST), Seoul 02792, South Korea}

\author{Hansol Kim}
\thanks{These authors contributed equally}

\affiliation{Center for Quantum Technology, Korea Institute of Science and Technology (KIST), Seoul 02792, South Korea}
\affiliation{Department of Quantum Information Science and Engineering, Sejong University, Seoul 05006, South Korea}

\author{Jaepil Park}
\affiliation{Center for Quantum Technology, Korea Institute of Science and Technology (KIST), Seoul 02792, South Korea}
\affiliation{Division of Quantum Information, KIST School, Korea University of Science and Technology, Seoul 02792, South Korea}

\author{Junhyung Lee}
\affiliation{Center for Quantum Technology, Korea Institute of Science and Technology (KIST), Seoul 02792, South Korea}
\affiliation{School of Electrical Engineering, Korea University, Seoul 02841, South Korea}

\author{Hye-yoon Jeon}
\affiliation{Center for Quantum Technology, Korea Institute of Science and Technology (KIST), Seoul 02792, South Korea}
\affiliation{School of Electrical Engineering, Korea University, Seoul 02841, South Korea}

\author{Jinil Lee}
\affiliation{Center for Quantum Technology, Korea Institute of Science and Technology (KIST), Seoul 02792, South Korea}
\affiliation{Division of Quantum Information, KIST School, Korea University of Science and Technology, Seoul 02792, South Korea}

\author{Yong-gwon Kim}
\affiliation{Center for Quantum Technology, Korea Institute of Science and Technology (KIST), Seoul 02792, South Korea}
\affiliation{Department of Physics, Kyung Hee University, Seoul 02447, South Korea}

\author{Yeeun Choi}
\affiliation{Center for Quantum Technology, Korea Institute of Science and Technology (KIST), Seoul 02792, South Korea}

\author{Min-Kyo Seo}
\affiliation{Department of Physics, Korea Advanced Institute of Science and Technology (KAIST), Daejeon, 34141, South Korea}

\author{Dae-Hwan Ahn}
\affiliation{Center for Quantum Technology, Korea Institute of Science and Technology (KIST), Seoul 02792, South Korea}
\affiliation{Division of Quantum Information, KIST School, Korea University of Science and Technology, Seoul 02792, South Korea}

\author{Hojoong Jung}
\affiliation{Center for Quantum Technology, Korea Institute of Science and Technology (KIST), Seoul 02792, South Korea}
\affiliation{Department of Physics, Kyung Hee University, Seoul 02447, South Korea}

\author{Dongyeon Daniel Kang}
\email{Co-Corresponding author: D.D.K.: dykang@kist.re.kr}
\affiliation{Center for Quantum Technology, Korea Institute of Science and Technology (KIST), Seoul 02792, South Korea}
\affiliation{Division of Quantum Information, KIST School, Korea University of Science and Technology, Seoul 02792, South Korea}

\author{Hyounghan Kwon}
\email{Co-Corresponding author: H.K.: hyounghankwon@kist.re.kr}
\affiliation{Center for Quantum Technology, Korea Institute of Science and Technology (KIST), Seoul 02792, South Korea}
\affiliation{Division of Quantum Information, KIST School, Korea University of Science and Technology, Seoul 02792, South Korea}

\maketitle

\textbf{Abstract.}
Integrated visible photonic engines for solid-state quantum defects provide a foundation for scalable quantum networks. While miniaturization is advancing, active manipulation remains limited by the difficulty of achieving simultaneous milliwatt-scale visible light generation and high-contrast modulation. Despite extensive efforts, the concurrent chip-scale realization of nonlinear frequency conversion and fast temporal gating for high-fidelity quantum control has remained elusive. Here, we demonstrate a monolithic thin-film lithium niobate (TFLN) platform integrating periodically poled frequency conversion with GHz-bandwidth electro-optic (EO) switching. The device delivers off-chip green-light power exceeding $1$~mW with an extinction ratio (ER) of $42.2$~dB, enabling coherent spin control and time-resolved lifetime measurements of individual nitrogen-vacancy (NV) centers in diamond through nanosecond gating. System performance is validated through pulsed optically detected magnetic resonance (ODMR), Rabi oscillations, and Ramsey interference, supported by time-tagged photon counting with nanosecond resolution. By unifying sufficient nonlinear light generation with high-speed active manipulation, this platform establishes a scalable framework for the realization of high-rate quantum communication nodes.





\section*{Introduction}
Realization of global quantum networks relies fundamentally on scalable light–matter interfaces where solid-state quantum defects serve as promising candidates for high-performance quantum memories and spin–photon nodes~\cite{duan2001long, kimble2008quantum}. These systems, comprising optically active quantum defects such as color centers in diamond~\cite{gruber1997scanning, dutt2007quantum} and silicon carbide~\cite{widmann2015coherent,bourassa2020entanglement,lukin20204h}, or rare-earth dopants in crystalline hosts~\cite{de2008solid, ruskuc2025multiplexed}, offer exceptional coherence properties and local processing capabilities~\cite{balasubramanian2009ultralong, maurer2012room}. The practical viability of these platforms has been underscored by recent milestones, including the demonstration of remote entanglement over long-distance fiber links and the realization of multi-node metropolitan networks~\cite{hensen2015loophole, hermans2022qubit, knaut2024entanglement}. These advances pave the way for transformative applications such as memory-enhanced quantum cryptography, blind quantum computing, and long-baseline quantum sensing~\cite{bhaskar2020experimental, wei2025universal, stas2025entanglement}. However, as these systems transition to high-rate multiplexed architectures which require multiple emitters within a single node, the complexity of the required optical control circuitry grows rapidly~\cite{lee2022quantum, wang2023field}. This reliance on bulky free-space implementations presents a scalability bottleneck, highlighting the need for integrated optical control layers to enable the large-scale realization of these quantum architectures~\cite{awschalom2018quantum, assumpcao2024thin, elshaari2020hybrid}.

Controlling these solid-state defects requires precise optical manipulation in the visible spectrum, which remains significantly more challenging than the mature telecommunications band~\cite{doherty2013nitrogen, sipahigil2016integrated, rugar2020generation, gottscholl2020initialization}. First, a primary obstacle lies in light generation itself, as the development of miniaturized, high-performance visible lasers is difficult due to material limitations such as the green gap~\cite{sun2024advancing, yuan2025efficient, pleasants2013overcoming}. Second, scaling these quantum systems requires independent optical addressing of multiple emitters~\cite{golter2023selective, cambria2025scalable, cai2021parallel}. Achieving this by multiplexing discrete conventional modulators introduces substantial system complexity and cost due to their bulky footprint. Furthermore, high-fidelity operation strictly necessitates both high-speed switching for time-resolved spin manipulation and high extinction contrast to suppress background noise~\cite{zhang2024scaled, zhao2025integrated}. In practice, these conventional devices impose a severe performance trade-off. Specifically, acousto-optic modulators (AOMs), the most widely adopted devices in current setups, provide the high extinction necessary to prevent photon leakage but are restricted to MHz-level bandwidths. Conversely, electro-optic (EO) modulators enable fast switching but suffer from low extinction ratio (ER).

Integrated visible photonic circuits provide a versatile and scalable framework for diverse optical technologies, with the control of solid-state quantum defects being a particularly prominent application~\cite{elshaari2020hybrid, majety2022quantum}. To this end, on-chip nonlinear frequency conversion has been extensively explored to circumvent the lack of native emitters spanning the ultraviolet to visible spectrum~\cite{wang2018ultrahigh, wu2024visible, hwang2023tunable}. In particular, substantial efforts have focused on developing on-chip green-light sources due to the green gap~\cite{sun2024advancing, yuan2025efficient, wang2025integrated, chen2025continuous}. Notably, silicon nitride-based devices have utilized the photogalvanic effect to induce effective second-order nonlinearities, successfully demonstrating on-chip green-light power reaching the milliwatt level~\cite{yuan2025efficient, wang2025integrated}. However, the centrosymmetric nature of silicon nitride precludes the intrinsic Pockels effect, necessitating complex heterogeneous integration for high-speed EO modulation~\cite{churaev2023heterogeneously, alexander2018nanophotonic}. In parallel, thin-film lithium niobate (TFLN) and thin-film lithium tantalate (TFLT) have received significant attention due to their inherently large second-order nonlinearities~\cite{wang2018ultrahigh, jankowski2020ultrabroadband, chen2025continuous}. For instance, TFLT platforms have successfully demonstrated milliwatt-level on-chip green power through quasi-phase matching (QPM) via periodic poling~\cite{chen2025continuous}. While TFLN-based modulator arrays have been employed to address solid-state quantum defects, these implementations still rely on bulky free-space optics, including external visible lasers and spatial light modulators~\cite{christen2025integrated}. Despite these great advances, the monolithic integration of efficient visible frequency generation and high-contrast, high-speed EO modulation within a single device, along with its direct application to the coherent manipulation of quantum defects, has not yet been demonstrated.

Here, we report a monolithic nonlinear EO visible photonic circuit based on a TFLN platform that addresses these long-standing challenges. Our device integrates a high-efficiency periodically poled lithium niobate (PPLN) frequency doubler with a GHz-bandwidth EO modulator on a single chip, demonstrating off-chip green power reaching the milliwatt level while enabling high-speed generation and precise gating of visible light. By exploiting the quadratic dependence of second-harmonic generation (SHG), the system fundamentally enhances the modulation contrast of the green light, achieving an ER exceeding 42 dB without the need for active feedback control. This nonlinear contrast enhancement represents a distinct operating regime that overcomes the performance trade-offs inherent in conventional linear EO modulators~\cite{hou2023programmable, kharel2021breaking}. As a proof-of-concept, we utilize this platform to demonstrate the optical addressing and coherent control of nitrogen-vacancy (NV) centers in diamond, showcasing the capability for high-fidelity manipulation across various temporal scales, including the nanosecond regime. Together, these results provide a viable pathway toward compact visible-light control hardware that can be scaled to support large-scale integrated quantum architectures.

\clearpage

\section*{Results}
\subsection*{Monolithic nonlinear EO visible photonics for quantum interfaces}

\begin{figure}
    \includegraphics[width=1\linewidth]{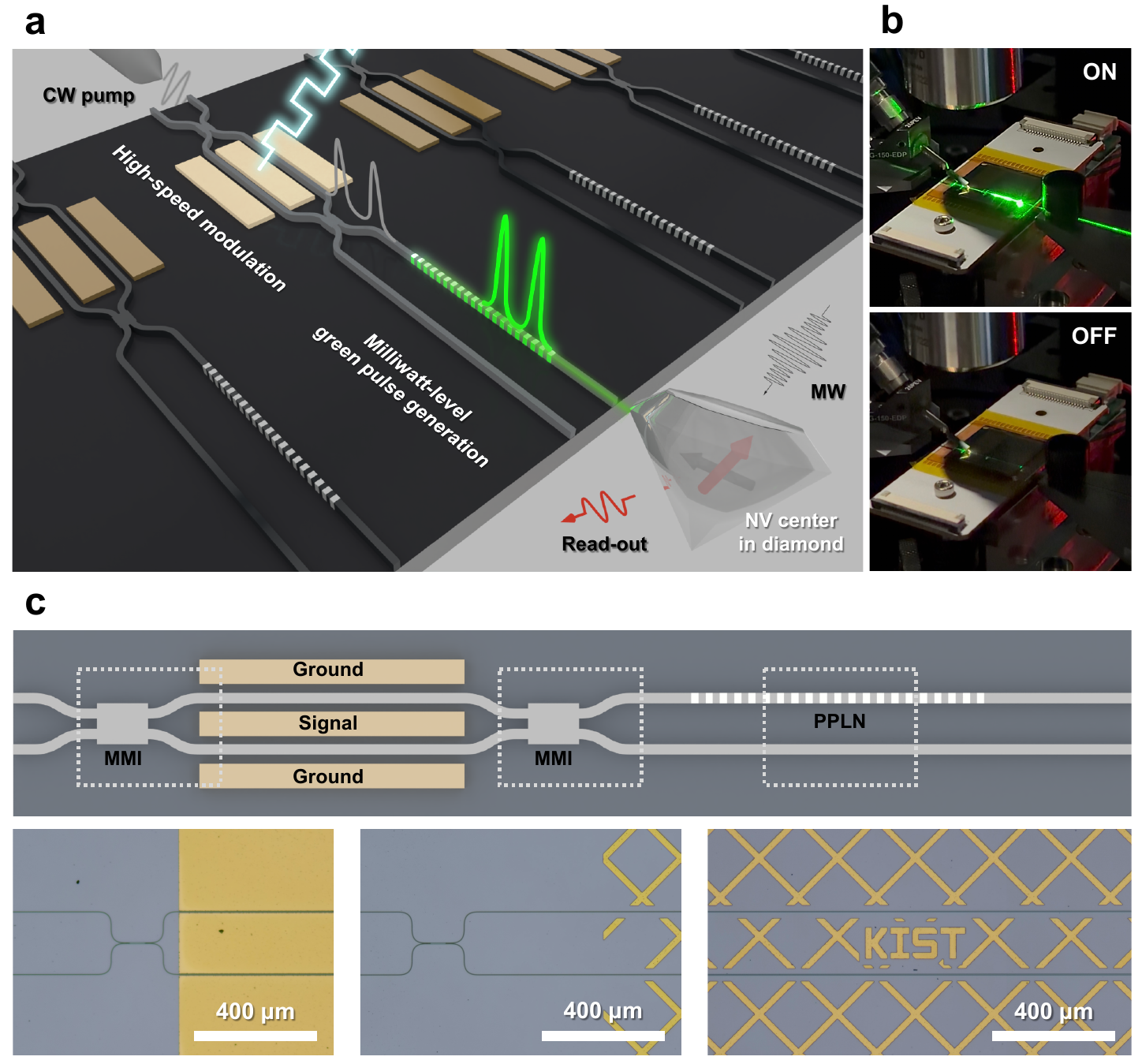}
    \caption{\label{fig:fig1}\textbf{Integrated thin-film lithium niobate nonlinear photonic circuits for high-speed green-light switching and optical addressing of an NV center in diamond.}  \textbf{a} Conceptual schematic of the integrated device and the quantum measurement. Continuous wave (CW) near-infrared pump light is modulated by an EO modulator and frequency-doubled to 532 nm in a PPLN waveguide, enabling milliwatt-level green light generation with fast on/off control and high extinction. The green light from the chip optically addresses an NV center in diamond for spin initialization and readout under microwave (MW) control. \textbf{b} Photographs of the fabricated device generating the green light guided on the chip in the ON (top) and OFF (bottom) states. \textbf{c} Top-view schematic layout of the device (top) and corresponding optical microscope images of the fabricated chip (bottom), showing the EO MZI switch and the PPLN waveguide sections.}
\end{figure}

Our integrated nonlinear EO visible photonic circuit, fabricated on a TFLN platform, enables high-speed and high-fidelity optical control of solid-state quantum defects. Figure~\ref{fig:fig1}a illustrates the conceptual schematic of the device alongside the overall optical addressing scheme for the NV centers in diamond. The monolithic circuit integrates an EO Mach-Zehnder interferometer (MZI) with a PPLN waveguide to achieve efficient SHG. System operation is initiated by coupling a 1064 nm CW near-infrared pump laser into the device, which is subsequently converted into a 532 nm green-light signal via a type-0 frequency doubling process.

The input pump light is first routed into the EO MZI section, which serves as a high-speed intensity modulator controlled by an external driving voltage. By leveraging the fast EO effect of TFLN, the MZI precisely shapes the temporal profile of the fundamental pump. The output power of the fundamental pump from the MZI, $P_{\omega}$, is governed by the interference and described by Equation~\ref{eq:mzi}:

\begin{equation}
    P_{\omega}(V) \propto \cos^2\left(\frac{\pi V}{2 V_{\pi}} + \phi_0\right)
    \label{eq:mzi}
\end{equation}
where $V$ is the applied drive voltage, $V_{\pi}$ is the half-wave voltage, and $\phi_0$ is the intrinsic phase bias. 

The modulated near-infrared signal subsequently propagates into the PPLN waveguide, where it is frequency-doubled to 532 nm via the $\chi^{(2)}$ nonlinear optical process. Crucially, by leveraging the quadratic dependence of the SHG process on the fundamental pump power, the generated green light power, $P_{2\omega}$, scales according to Equation~\ref{eq:shg}:

\begin{equation}
    P_{2\omega}(V) \propto \left[ P_{\omega}(V) \right]^2 \propto \cos^4\left(\frac{\pi V}{2 V_{\pi}} + \phi_0\right)
    \label{eq:shg}
\end{equation}
This quadratic relation effectively squares the ER of the fundamental pump. Consequently, this nonlinear gating mechanism overcomes the intrinsic extinction limits of standard EO modulators without requiring complex feedback systems, achieving an exceptional ER exceeding 40 dB. We note that although the input pump is carefully aligned to maximize the TE component, the achievable TE polarization may be limited by the polarization extinction of the input fiber. In this regard, the type-0 phase-matching condition enabling TE SH mode generation from the TE pump inherently suppresses residual non-TE components, allowing the SHG process to act as an additional polarization filter. As demonstrated in Fig.~\ref{fig:fig1}b, the device successfully generates milliwatt-level guided green light, with the macroscopic ON and OFF states exhibiting highly visible contrast.

Figure~\ref{fig:fig1}c details the top-view schematic layout alongside corresponding optical microscope images of the fabricated monolithic circuit (See Methods for details on fabrication). These images display the cascaded configuration of the device, specifically showing the MZI switch and the subsequent PPLN waveguide section. Notably, the PPLN region incorporates an Au mesh structure designed to suppress parasitic optical leakage into the slab mode~\cite{stokowski2023integrated}. Building upon the high-contrast switching of this integrated platform, the generated visible light is routed off-chip to optically address an individual NV center in diamond, facilitating spin initialization and time-resolved readout.

\clearpage

\subsection*{Efficient green light generation through SHG in TFLN waveguides}

\begin{figure}
    \includegraphics[width=1\linewidth]{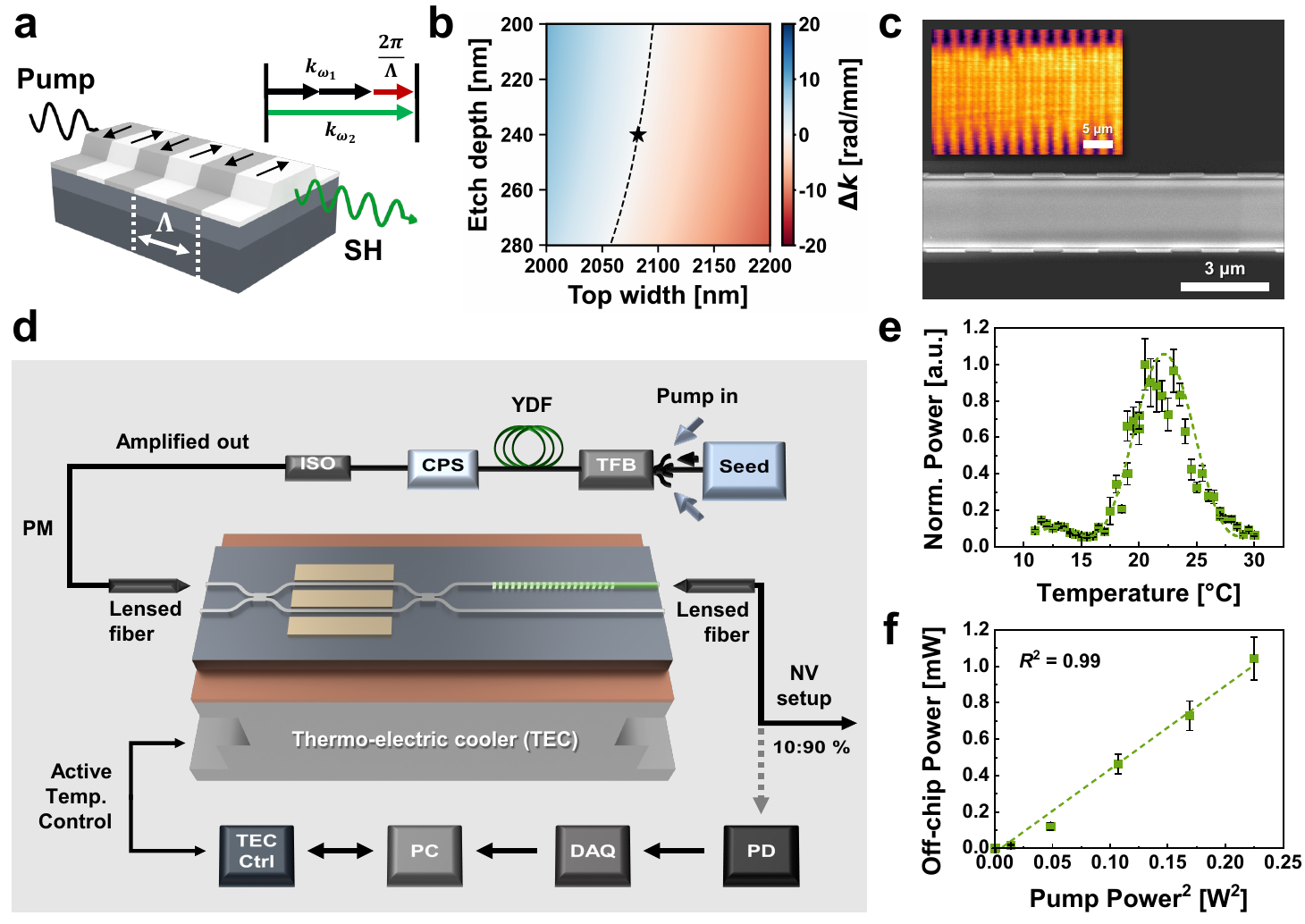}
    \caption{\label{fig:fig2} \textbf{Efficient green light generation in a PPLN waveguide.} \textbf{a} Conceptual schematic of the PPLN section, where the near-infrared fundamental pump light is converted to the green light via domain-inverted QPM. $\Lambda$ denotes the poling period. \textbf{b} Calculated phase mismatch ($\Delta k$) map versus waveguide top width and etch depth. The marked point indicates the designed geometry used for the device. \textbf{c} Scanning electron microscopy (SEM) image of the fabricated PPLN waveguide (scale bar: 3 $\mu$m). Inset shows the SHG microscopy image of the domain pattern before waveguide fabrication (scale bar: 5 $\mu$m). \textbf{d} Schematic of the experimental setup utilizing a fiber-amplified pump source and lensed fiber coupling. Temperature stabilization via a TEC is applied to maintain the optimal QPM condition. Active fiber alignment stabilization is also performed by monitoring the output power. \textbf{e} Temperature-dependent SHG response showing the QPM tuning curve. The dashed line is a sinc$^2$ fit used to identify the optimum phase-matching temperature. \textbf{f} Off-chip measured SHG power as a function of the squared pump power. The green dashed line shows the expected quadratic scaling, corresponding to a normalized conversion efficiency of $\sim$0.46\%/W.}    
\end{figure}

We evaluate the SHG performance of the integrated PPLN stage to validate its frequency-conversion efficiency and phase-matching characteristics. As shown in Fig.~\ref{fig:fig2}a, efficient on-chip green-light generation is enabled by QPM in a PPLN ridge waveguide with a poling period of $\Lambda$. The periodic poling introduces a reciprocal lattice vector of $2\pi/\Lambda$ that compensates the modal phase mismatch $\Delta k$, satisfying
\begin{equation}
\Delta k = k_{\mathrm{SH}} - 2k_{\mathrm{Pump}} - \frac{2\pi}{\Lambda}=0,
\end{equation}
where $k_{\mathrm{Pump}}$ and $k_{\mathrm{SH}}$ denote the wavevectors of the pump and second-harmonic (SH) waves, respectively. The required poling period $\Lambda$ can be expressed in terms of the effective indices as
\begin{equation}
\Lambda = \frac{\lambda_{\mathrm{SH}}}{n_{\mathrm{SH}} - n_{\mathrm{Pump}}},
\label{eq:qpm_period}
\end{equation}
where $\lambda_{\mathrm{SH}}$ is the SH wavelength and $n_{\mathrm{Pump}}$ and $n_{\mathrm{SH}}$ are the effective indices of the guided pump and SH modes, respectively.

We design the PPLN ridge waveguide to support Type-0 interaction and to ensure fundamental-mode operation at both wavelengths. This mode-selective design provides an output profile directly compatible with single-mode fiber (SMF) coupling for subsequent NV experiments. For our X-cut TFLN device, the TE fundamental modes are engineered to be consistent with the Type-0 polarization configuration, enabling access to the largest nonlinear coefficient, $d_{33}$. To determine the optimal parameters, we perform finite-difference eigenmode (FDE) simulations to compute the effective indices and map the residual phase mismatch over a two-dimensional sweep of ridge top width and etch depth, as illustrated in Fig.~\ref{fig:fig2}b. This map identifies the zero-mismatch contour and the sensitivity of QPM to dimensional variations. From this analysis, we select a poling period of $\Lambda = 2.25~\mu\text{m}$ and an optimized ridge geometry with a top width of $2082$~nm and an etch depth of $240$~nm, which is marked as a black star in Fig.~\ref{fig:fig2}b, to ensure QPM at the target wavelengths.

The structural and nonlinear integrity of the fabricated PPLN device is characterized using scanning electron and nonlinear optical microscopy, as presented in Fig.~\ref{fig:fig2}c. Top-view SEM imaging reveals periodic sidewall grooves resulting from domain-dependent anisotropic wet etching, which provides a direct confirmation of the targeted poling period along the waveguide. Furthermore, the uniformity of the domain inversion is validated via the SHG microscopy image in the inset of Fig.~\ref{fig:fig2}\textbf{c}, where the consistent signal contrast across alternating domains indicates a homogeneous nonlinear response throughout the poled region.

As illustrated in Fig.~\ref{fig:fig2}d, the SHG performance of the integrated device is characterized using an in-house 1064~nm fiber laser system, coupled to the chip via lensed fibers and piezo-actuated stages. To facilitate efficient frequency conversion, the seed signal is boosted by a custom Ytterbium-doped fiber amplifier (YDFA), providing an on-chip pump power of several hundred milliwatts. To maintain the optimal QPM condition and mitigate environmental drifts, the setup incorporates active temperature regulation via a thermoelectric cooler (TEC) alongside a real-time fiber-to-chip alignment feedback loop. This stabilization framework ensures a consistent nonlinear response by minimizing photothermal and mechanical fluctuations throughout the measurement process. (See Methods for details of the experimental configuration.)

We evaluate the SHG performance of the integrated PPLN waveguide to verify its suitability as a reliable excitation source providing sufficient optical power for the initialization and readout of solid-state spin defects. To define the optimal operational condition, we first mapped the thermal landscape of the QPM condition by monitoring the SHG signal across a range of device temperatures. As illustrated in Fig.~\ref{fig:fig2}e, the temperature-dependent power tuning curve broadly follows the characteristic $\text{sinc}^2$ profile, with a numerical fit identifying the peak phase-matching temperature near room temperature. 

Following the identification of this optimal operating point, we investigated the output power scaling as a function of the fundamental pump power, as shown in Fig.~\ref{fig:fig2}f. The generated green light exhibits a clear quadratic dependence on the pump power, confirming that the device operates within the predicted nonlinear regime. Notably, the measured off-chip green output exceeds $1$~mW, which is already sufficient for high-fidelity initialization and readout of NV center spins. To provide a reference for the system's coupling quality, we measured a total fiber-to-fiber loss of $14.3$~dB at the pump wavelength through the MZI. While the internal green-light power is strictly higher due to these coupling and propagation losses, we adopt the measured off-chip value as the primary and conservative benchmark for the system's performance. By focusing on measured values, we avoid the uncertainties of modeling internal losses, providing a conservative yet sufficient performance metric for high-fidelity quantum addressing.


\clearpage

\subsection*{High-extinction EO modulation and green-light pulse gating}
\begin{figure}
    \includegraphics[width=1\linewidth]{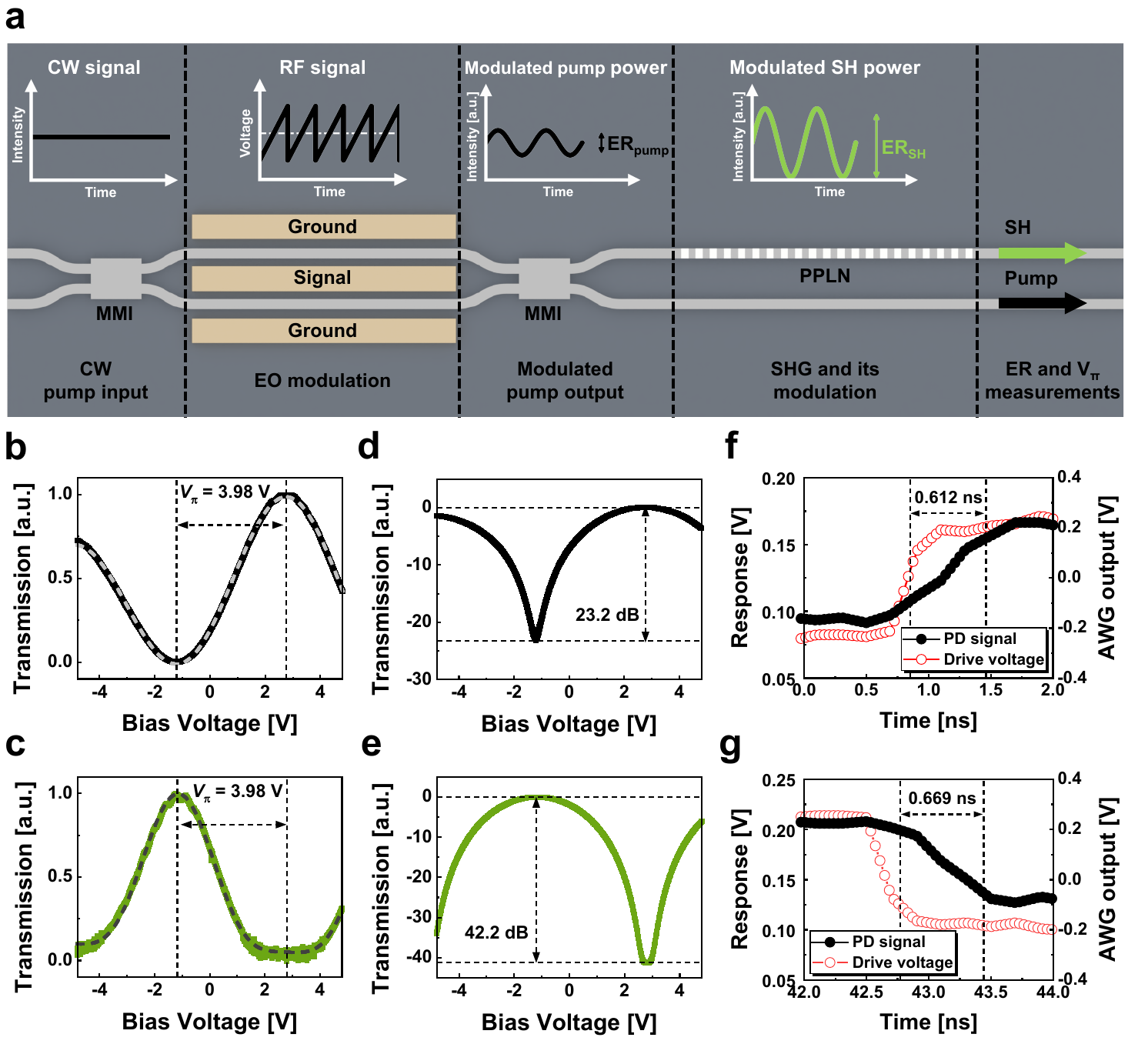}
    \caption{\label{fig:fig3} \textbf{Dynamic modulation of the fundamental (1064 nm) and second-harmonic (532 nm) signals.}
    \textbf{a} Operational principle and measurement flow. The 1064 nm pump light is intensity-modulated by an EO MZI and frequency-doubled to 532 nm in the PPLN section. This allows for simultaneous characterization of the extinction ratios for the pump ($ER_{pump}$) and the second-harmonic ($ER_{SH}$) signals. \textbf{b, c} Normalized transmission versus DC bias voltage. This yields $V_{\pi} = 3.98$ V for 1064 nm and $V_{\pi} = 3.98$ V for 532 nm. Dashed curves indicate model fits reflecting the sinusoidal MZI response and quadratic scaling of SH. \textbf{d, e} Transmission in logarithmic scale. The data demonstrate extinction ratios of 23.2 dB for the pump and 42.2 dB for the SH signal. \textbf{f, g} Time-domain switching measured with a photodetector. Measured $10\text{--}90\%$ rise and fall transition times are $0.612~\text{ns}$ and $0.669~\text{ns}$, respectively, approaching the $1.2~\text{GHz}$ bandwidth limit of the photodetector.}
\end{figure}

Building on the robust generation of milliwatt-scale green light, we characterize the active modulation capabilities of our platform for high-speed quantum addressing. As illustrated in Fig.~\ref{fig:fig3}a, the 1064 nm pump is intensity-modulated using an on-chip EO MZI and subsequently frequency-doubled to 532 nm in the PPLN section, allowing simultaneous evaluation of the extinction behavior at both the fundamental and second-harmonic wavelengths.

We first evaluate the DC transfer characteristics by measuring the transmission response as a function of the applied bias voltage. Figures~\ref{fig:fig3}b and \ref{fig:fig3}c show the normalized transmission for the pump and SH signals, respectively, both exhibiting the expected sinusoidal modulation behavior. The dashed curves represent model fits reflecting the sinusoidal pump response and the quadratic scaling of the SH signal. From these measurements, the half-wave voltage is extracted to be $V_{\pi}=3.98$ V for both wavelengths. This consistent $V_{\pi}$ confirms stable EO phase control and robust operation of the integrated MZI across the nonlinear conversion process.

The modulation contrast is quantified by the extinction ratios of the fundamental pump ($ER_{\text{pump}}$) and the generated second-harmonic ($ER_{\text{SH}}$) signals, as determined from the logarithmic transmission data. Figures~\ref{fig:fig3}d and~\ref{fig:fig3}e illustrate these metrics, where the pump signal exhibits an $ER_{\text{pump}}$ of 23.2~dB, while the SH output achieves an enhanced $ER_{\text{SH}}$ of 42.2~dB. This substantial improvement originates from the quadratic dependence of the SHG process on the pump field. In the ideal case, $ER_{\text{SH}}$ in decibels is expected to be double that of $ER_{\text{pump}}$. In our characterization, the observed enhancement factor of 1.819 is slightly below this theoretical limit of 2. However, the minimum transmission levels shown in Figures~\ref{fig:fig3}e are clearly limited by the photodetector noise floor, implying that the intrinsic $ER_{\text{SH}}$ is likely even higher and could be further revealed with increased signal-to-noise ratios. It is worth noting that such high-contrast modulation is achieved through a single-stage nonlinear process, bypassing the complexity of cascaded modulators or active feedback control.

The dynamic switching performance is further evaluated through time-domain measurements, as presented in Figures~\ref{fig:fig3}f and ~\ref{fig:fig3}g. The modulated optical pulses closely follow the applied RF drive, exhibiting transition rise and fall times of approximately 0.612~ns and 0.669~ns, respectively. These response times approach approach the $1.2~\text{GHz}$ bandwidth of the photodetector (Thorlabs DET01CFC), noting the potential for even faster intrinsic switching speeds beyond these measurement results. This sub-nanosecond gating capability, combined with the high extinction ratio, confirms the platform's suitability for time-resolved optical addressing of solid-state quantum defects.

\clearpage
\subsection*{Optical addressing and CW-ODMR spectroscopy of NV centers}
\begin{figure}
    \includegraphics[width=1\linewidth]{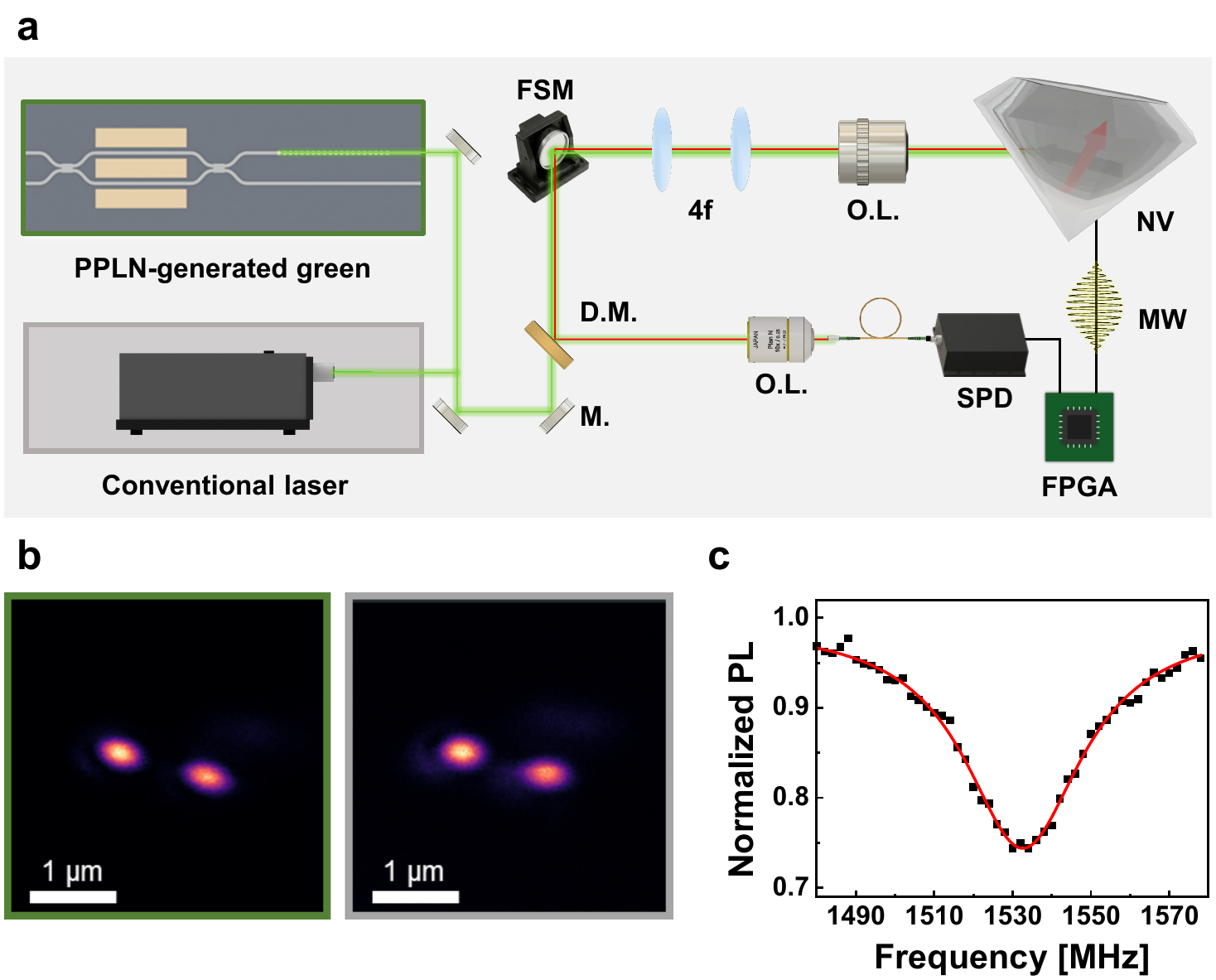}
    \caption{\label{fig:fig4} \textbf{Photonic chip-based confocal imaging and CW-ODMR measurements of NV centers.} \textbf{a} Schematic of the experimental setup. A 1064 nm pump laser is frequency-converted in the PPLN nonlinear device to generate 532 nm excitation light. This is delivered to a confocal microscopy setup for NV excitation and fluorescence collection. The performance is benchmarked against a conventional 532 nm laser under the same optical configuration (green box: PPLN-generated 532 nm source; grey box: conventional 532 nm laser). \textbf{b} Confocal fluorescence images of NV centers acquired under identical conditions. \textbf{c} CW-ODMR spectrum measured using the on-chip generated 532 nm excitation. A resonance dip in fluorescence is observed as the microwave frequency is scanned across the spin transition.}
\end{figure}

We validate the direct deployment of the chip-generated green light for optical addressing of NV centers in diamond. In these systems, green light excitation serves as the primary source for both non-resonant excitation and spin-state initialization into the ground state. We provide the CW initialization and spin-dependent readout required for ODMR spectroscopy without any modification to the standard confocal microscope setup (see Methods for details in measurement and sample preparation). The experimental configuration, as illustrated in Fig.~\ref{fig:fig4}a, was utilized to benchmark the PPLN frequency-converted output by fiber-delivering the chip-generated green light into the existing setup for a direct comparison against a conventional bulk 532~nm laser. This approach ensures that both excitation sources share an identical optical path and detection setup, allowing for a rigorous and systematic performance evaluation.

As shown in the confocal maps in Fig.~\ref{fig:fig4}b, the green light from the chip enables high-contrast imaging and precise localization of individual emitters with diffraction-limited resolution (see Methods for details in the confocal imaging). Furthermore, this TFLN-based source provides the necessary power stability to maintain a consistent signal-to-noise ratio during extended raster scans, yielding results comparable to conventional bulk lasers in Fig.~\ref{fig:fig4}b. Such performance validates the platform as a reliable excitation source for high-resolution single-emitter microscopy.

We then demonstrate the spin-state initialization and readout capabilities of the green light from the chip through CW-ODMR spectroscopy. Beyond simple fluorescence excitation, high-fidelity quantum control of NV centers requires effective optical pumping into the ground state and subsequent spin-dependent contrast for readout. As shown in the resonance spectra in Fig.~\ref{fig:fig4}c, the chip-generated excitation successfully supports these essential spin-physics processes, yielding a pronounced ODMR dip with a fluorescence contrast of $\sim$25\% and a linewidth of 36~MHz. These results are in excellent agreement with the benchmarks obtained from the conventional bulk 532~nm laser, confirming that the TFLN platform is fully compatible with high-contrast spin spectroscopy. The TFLN engine thus provides a viable integrated alternative to bulk visible excitation for standard quantum-defect experiments.

\clearpage

\clearpage

\subsection*{High-speed pulse gating for coherent spin manipulation}
\begin{figure}
    \includegraphics[width=1\linewidth]{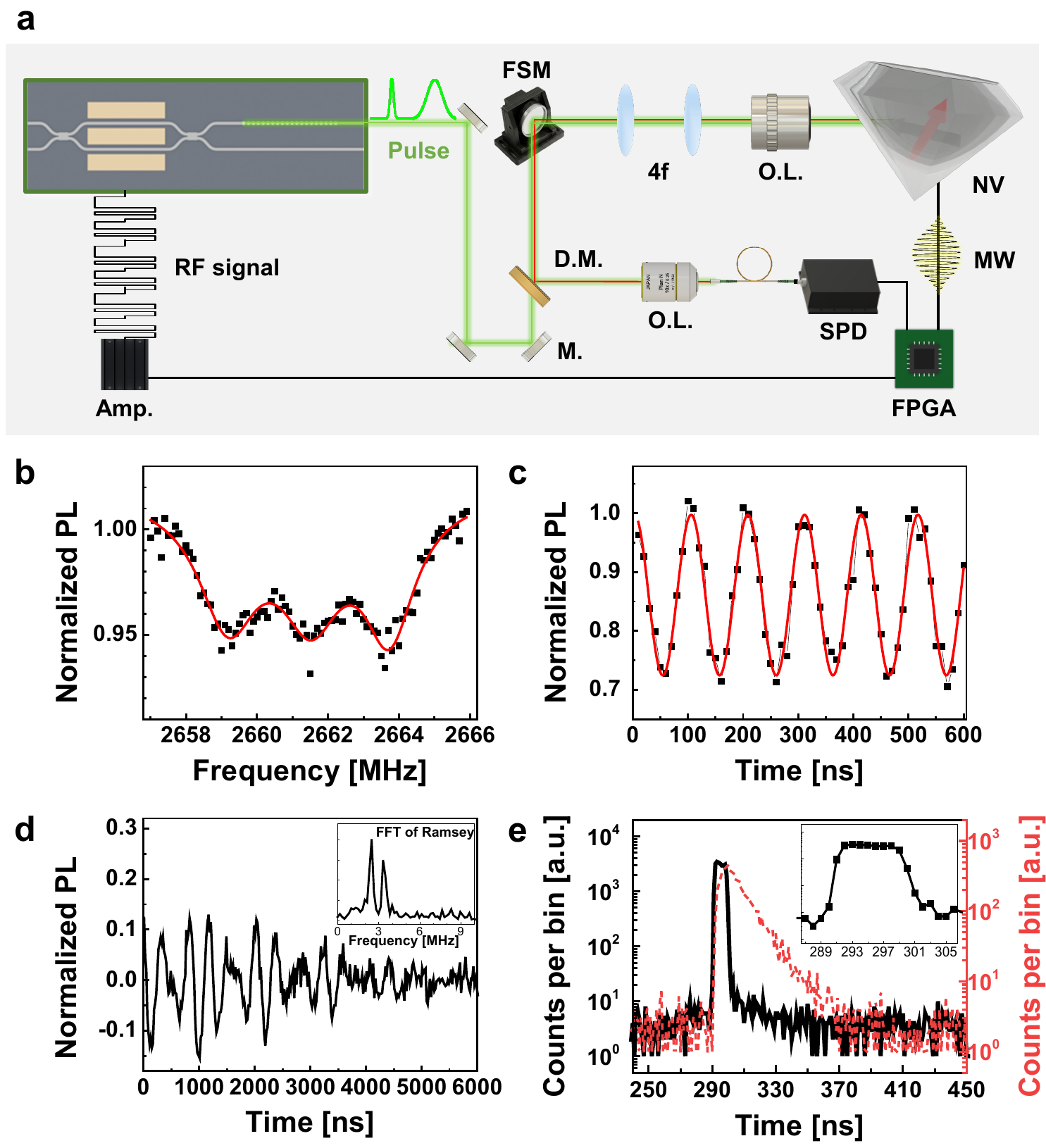}
    \caption{\label{fig:fig5} \textbf{High-speed EO pulse gating for coherent spin measurements of NV centers.}
    \textbf{a} Schematic of the FPGA-synchronized pulse control scheme. Voltage-driven PPLN switching enables nanosecond-scale optical pulse delivery to NV centers combined with time-tagged photon counting. This configuration facilitates pulsed ODMR, Rabi oscillations, Ramsey interference, and time-resolved fluorescence measurements. \textbf{b} Pulsed ODMR spectrum of a single NV center. Multiple resonance dips are resolved as the microwave frequency is scanned. \textbf{c} Rabi oscillations obtained by varying the microwave pulse duration. \textbf{d} Ramsey interference fringes. These are used to extract the spin dephasing time $T_2^*$ and detuning-dependent phase evolution. The inset shows the FFT of the Ramsey trace highlighting dominant frequency components. \textbf{e} Time-resolved fluorescence trace following pulsed excitation. The black line indicates the PPLN-generated optical pulse profile , while the red dashed line shows the NV fluorescence decay triggered by the pulse. The inset displays a magnified view of the PPLN-generated green light pulse.}
\end{figure}

Following the CW validation, we demonstrate the integrated EO modulation stage for coherent spin control, providing both high-speed gating and high-contrast modulation. Figure~\ref{fig:fig5}a illustrates the optical setup, where the TFLN circuit is electrically driven by field-programmable gate array (FPGA)-synchronized voltage waveforms to generate user-defined 532~nm pulse patterns directly on-chip. As shown in Figs~\ref{fig:fig3}e--\ref{fig:fig3}g, the TFLN photonic circuits enable sub-nanosecond switching transitions alongside a high extinction ratio of 42.2~dB, ensuring the timing fidelity and contrast required for pulsed quantum protocols. These on-chip pulses are hardware-synchronized with MW driving and photon detection, providing the optical control interface for high-fidelity electron spin manipulation of NV center.

To evaluate the performance of the chip-based high-speed pulse gating for reliable spin control, we conduct a series of coherent spin-manipulation measurements. We first verify the time-gated optical readout through the pulsed ODMR measurements shown in Fig.~\ref{fig:fig5}b. The resolved triplet structure from the host $^{14}\text{N}$ nuclear spin confirms that on-chip EO gating provides sufficient extinction and timing for reliable initialization and gated fluorescence readout. Coherent spin manipulation is further established by the Rabi oscillations presented in Fig.~\ref{fig:fig5}c. The observed 51~ns period allows for precise calibration of microwave pulses, while the high Rabi contrast sustained over five periods demonstrates that residual off-state leakage is negligible thanks to the high $ER_{\text{SH}}$. The phase integrity of the experiment is further validated through Ramsey interferometry with detuning, as shown in Fig.~\ref{fig:fig5}d. The clear Ramsey fringes and the resolved $^{13}\text{C}$ nuclear spin splittings in the fast Fourier transform (FFT) indicate that the FPGA-synchronized green pulses maintain the timing stability and high contrast necessary for free evolution.

To highlight the advantages of our integrated approach over the speed of conventional AOMs which typically needs tens of nanoseconds for rise/fall, we time-resolve the chip-generated green pulse and the subsequent NV fluorescence decay in Fig.~\ref{fig:fig5}e. The black trace, measured from the surface-reflected excitation pulse, and the red dashed trace, showing the NV fluorescence decay, reveal clear temporal separation with sub-nanosecond pulse definition. Although the off-state baseline is set by the APD dark counts, the high on/off contrast is maintained under the same conditions used for coherent spin control. These results demonstrate that our platform provides a compact solution that simultaneously achieves high-speed switching and high extinction, overcoming the respective limitations of AOMs and bulk EOMs in the visible regime. Collectively, our on-chip nonlinear EO visible photonic circuit serves as a functional and scalable replacement for conventional pulse pickers in standard solid-state quantum experiments.

\clearpage

\section*{Discussion and outlook}

In this study, we have demonstrated a monolithic TFLN platform integrating high-efficiency SHG and GHz-bandwidth EO switching to provide a high-performance control interface for quantum systems. The architecture delivers off-chip visible-light power exceeding 1~mW with a high $ER_{\text{SH}}$ of 42.2~dB, which matches the performance standards of bench-top free-space systems. By utilizing nanosecond-scale on-chip gating, we achieved high-fidelity coherent spin control of individual NV centers, as evidenced by clear Rabi oscillations and Ramsey interference. Notably, our time-tagged photon counting with nanosecond resolution directly resolves the on-chip green pulses and the subsequent NV fluorescence decay, providing a definitive physical validation of the gating performance. In addition to the high-speed performance, these results validate that the integrated device maintains the temporal stability required for extended quantum measurements, while suggesting that a monolithic architecture can simplify the alignment overhead within a compact footprint.

Integrated green light generation through nonlinear processes has been extensively explored across various material platforms with recent notable achievements exceeding $1$~mW on chip summarized in Table 1~\cite{yuan2025efficient, wang2025integrated, chen2025continuous}. While these previous works demonstrate impressive power levels, the simultaneous realization of high output power and high-speed modulation, which is essential for active quantum control, remains a significant challenge for most existing works. Our monolithic TFLN platform addresses this gap by delivering competitive off-chip power alongside GHz-bandwidth EO switching. The availability of milliwatt-level net power ensures that sufficient excitation intensity can be delivered to emitter sites with sub-nanosecond precision, meeting the stringent technical requirements for the scalable control of solid-state quantum defects within an integrated format. Furthermore, the achievement of the milliwatt-level green power can extend the utility of this platform beyond quantum science to diverse fields, including biomedical imaging, high-speed visible light communication, and compact light engines for next-generation display technology.

The architecture presented here offers a versatile framework for expansion across spectral, spatial, and system-level domains. By leveraging the flexibility of periodic poling design, the platform can be tailored to generate various wavelengths across the visible spectrum using readily available infrared lasers~\cite{wu2024visible,jankowski2020ultrabroadband}. This spectral versatility facilitates a universal control interface for a diverse range of solid-state quantum emitters, such as color centers in silicon carbide or rare-earth ions, extending its utility beyond the specific requirements of NV centers~\cite{christle2015isolated,de2008solid,luo2023recent}. Furthermore, the compact footprint is well-suited for scaling into high-density modulator arrays~\cite{assumpcao2024thin,christen2025integrated}, which could enable independent and simultaneous addressing of multiple emitters with reduced spatial constraints. In parallel, the integration of permanent fiber-to-chip packaging is expected to provide the long-term stability required for practical quantum implementations~\cite{weninger2026advances}.

Looking forward, the development of large-scale quantum information systems will likely benefit from overcoming the spatial and operational constraints of conventional bulk optical architectures. The integration of frequency conversion and high-speed modulation within a monolithic format provides a practical route to mitigate scalability challenges in the visible spectrum while addressing the performance limitations that have long constrained quantum control. This interface offers a versatile framework for the independent manipulation of diverse solid-state emitters, potentially simplifying the operational requirements for large-scale, interconnected nodes. Such an integrated architecture represents a significant step toward the deployment of high-rate quantum nodes, contributing to the development of robust, large-scale quantum communication networks.

\clearpage

\section*{Methods}
\subsection*{Device design and fabrication}
The effective mode indices are simulated using commercially available software, Ansys Lumerical MODE.

The device fabrication consists of two primary stages, beginning with ferroelectric domain inversion prior to waveguide definition. The 300-nm-thick TFLN film was first cleaned, followed by spin coating of Poly(methyl methacrylate) (PMMA 950 A5) at 2000 rpm. To mitigate electron-beam charging during lithography, an additional conductive e-spacer layer was deposited on top of the resist stack. The poling electrode pattern was then defined using electron-beam lithography (JEOL JBX-9300FS). After development in a mixed solution of deionized water and isopropyl alcohol, a 100-nm chromium layer was deposited by electron-beam evaporation and subsequently lifted off in acetone. Periodic ferroelectric domain inversion for QPM was achieved by applying high-voltage electrical pulses across the patterned electrodes. Following the poling process, the chromium electrodes were removed using a chromium etchant, while the alignment markers were preserved for subsequent fabrication steps.

The subsequent fabrication stage focuses on waveguide definition. Hydrogen silsesquioxane (HSQ) resist was spin-coated at 2000 rpm, followed by the deposition of an e-spacer layer to suppress charging effects during electron-beam exposure. The waveguide geometry was patterned using the same electron-beam lithography system. After development in AZ300MIF, the structures were transferred into the TFLN layer using inductively coupled plasma reactive ion etching (ICP–RIE) with Ar ions. The residual resist was then removed using buffered oxide etchant (BOE), followed by a KOH cleaning step. A SiO$_2$ upper cladding was subsequently deposited by plasma-enhanced chemical vapor deposition (PECVD), and the fabrication was completed with a final annealing process.

\subsection*{Measurement of on-chip nonlinear visible light generation in the TFLN platform}
The SHG measurements in Fig.~\ref{fig:fig2} were performed using a custom-built 1064~nm fiber laser system. The seed signal was amplified by launching it into the core of a 4-m-long double-clad ytterbium-doped fiber (YDF, Coherent-Nufern PLMA-YDF-10/125-M) via a tapered fiber bundle (TFB, AFR PMMPC-2+1X1-06-976-22-SS-SS-1-1). A 976~nm diode laser was coupled into the YDF cladding to provide the necessary pump power. To ensure signal purity and system safety, residual pump light was removed using a cladding power stripper (CPS, AFR PMCPS-976-06-10-020-2-1), and a high-power optical isolator (ISO) was integrated to suppress back-reflections. The resulting YDFA output provided a tunable pump power range of 0.5-1~W. All optical paths were implemented using polarization-maintaining (PM) fibers, with the input polarization precisely aligned to the fundamental TE mode of the TFLN device. For thermal management, the chip was mounted on a copper heat sink controlled by a thermoelectric cooler (TEC) to maintain the optimal QPM temperature. Furthermore, an active alignment system was employed, utilizing a real-time feedback loop that monitored the output power to compensate for mechanical drifts, thereby ensuring long-term coupling stability throughout the characterization process.

\subsection*{Experimental configuration for spin control of NV center in diamond}

NV centers in diamond (quantum-grade from Element Six), were created by electron-beam irradiation followed by annealing at $1100\,^{\circ}\mathrm{C}$. Moreover, solid-immersion lens structures were fabricated on the diamond surface, and gold microwave delivery patterns were additionally fabricated in close proximity to the solid-immersion lens to enable efficient spin control of NV centres hosted within the solid-immersion lens.

Measurements were performed at room temperature using a custom-built beam-scanning confocal microscope system. To enable high-speed raster scanning, the setup was equipped with a fast steering mirror (FSM) and a 4f relay system to conjugate the scanning pivot to the objective's back focal plane. Individual NV centers in diamond were excited at $532~\mathrm{nm}$, and their fluorescence were collected using a Zeiss 100$\times$ air objective ($\text{NA} = 0.95$) and coupled into a step-index multimode fiber through a 10$\times$ objective (Thorlabs RMS10X), where the fiber core acted as the confocal aperture. Single photons were detected via a fiber-coupled silicon avalanche photodiode (APD, Excelitas Technologies, SPCM-AQRH-14-FC). This enabled high-contrast imaging and precise localization of individual emitters with a diffraction-limited lateral resolution of $\sim 0.3~\mu\mathrm{m}$ (Rayleigh criterion, $0.61\lambda/\mathrm{NA}$) in the case without solid-immersion lens.

Microwave signals were generated and timing-programmed by an FPGA-based controller (AMD Zynq UltraScale+ RFSoC ZCU111) and then amplified by a power amplifier. The microwave field was delivered to near-surface NV centres located in a solid-immersion lens. A static magnetic field (74 G or 496 G) was applied using a permanent magnet to lift the spin-state degeneracy. For all spin measurements, the microwave control (including frequency sweeps and pulse sequences) and photon integration were synchronized by the FPGA-based timing controller to ensure stable spectra and reproducible time-domain signals.

\section*{Acknowledgments}
This work was supported by National Research Foundation (NRF) (RS-2023-NR119925,RS-2024-00343768,RS-2024-00509800,RS-2025-25445839), National Research Council of Science and Technology (NST) (GTL25011-000), Institute for Information \& communications Technology Planning \& Evaluation (IITP) grant funded by the Korea government (MSIT) (RS-2025-02218723,RS-2025-25464657), and Korea Institute of Science and Technology (KIST) research program (26E0001,26E0011). The authors acknowledged the support by National Information Society Agency(NIA) funded by the Ministry of Science and ICT(MSIT, Korea) [Quantum technology test verification and consulting support in 2024].

\section*{Conflict of interests}
\noindent The authors declare no competing financial interests.

\section*{Author contributions}
\noindent Y.P., Y.S.L., and Hansol Kim contributed equally to this work. Y.P. and Hansol Kim designed the devices. Y.P., Junhyung Lee, Hye-yoon Jeon, and Jinil Lee fabricated the devices. Hansol Kim built the in-house Ytterbium fiber laser. Y.P. and Hansol Kim performed the characterization of the TFLN photonic circuits with assistance from Junhyung Lee. Y.C. fabricated the diamond sample. Y.S.L., Y.P., and Hansol Kim performed the chip-based NV-center experiments with assistance from J.P. and Y.-G.K. Hyounghan Kwon and D.D.K. conceived and supervised the project. Y.P., Y.S.L., Hansol Kim, D.D.K., and Hyounghan Kwon wrote the manuscript. All authors discussed the results and commented on the manuscript.

\section*{Data Availability Statement}
\noindent The data that support the findings of this study are available from the corresponding author upon request.

\clearpage
\section{References}
\bibliographystyle{naturemag_noURL}
\bibliography{reference}
\clearpage

\clearpage

\clearpage

\clearpage

\clearpage

\begin{table}[t]
\centering
\caption{\textbf{Benchmarking of milliwatt-scale integrated nonlinear green-light sources.}}
\label{tab:comparison_visible_sources}
\footnotesize
\setlength{\tabcolsep}{0pt}
\renewcommand{\arraystretch}{1.3}

\begin{tabular*}{\linewidth}{@{\extracolsep{\fill}}lccccc@{}}
\toprule
\multicolumn{1}{c}{Platform} & Process & \makecell{Phase-matching\\condition} & \makecell{Green light\\power} & \makecell{Switch\\integration} & \makecell{Quantum\\application} \\
\midrule
TFLT~\cite{chen2025continuous} & SHG & QPM & 1.87\,mW (On-chip) & \textsf{X} & \textsf{X} \\
SiN~\cite{wang2025integrated}  & SHG & Induced $\chi^{(2)}$ & 3.5\,mW (On-chip) & \textsf{X} & \textsf{X} \\
SiN~\cite{yuan2025efficient}   & SHG & Induced $\chi^{(2)}$ & 5.3\,mW (On-chip) & \textsf{X} & \textsf{X} \\
\textbf{TFLN (This work)} & \textbf{SHG} & \textbf{QPM} & \textbf{1\,mW (Off-chip)} & \textbf{EO (GHz-level)} & \textbf{NV spin addressing} \\
\bottomrule
\end{tabular*}
\end{table}



\end{document}